\documentclass[jcp,twocolumn]{revtex4-1}
\usepackage{amsmath}
\usepackage[english]{babel}
\usepackage{latexsym}
\usepackage{mathrsfs}
\usepackage{graphicx}
\usepackage{graphics}
\usepackage{color}
\usepackage{bm}
\usepackage{units,nicefrac}

\begin{document}

\title{Amorphous Ni$_{50}$Ti$_{50}$ Alloy with Nanoporous Structure \\ Generated by Ultrafast Isobaric Cooling}

\author{Bulat N. Galimzyanov} \email{bulatgnmail@gmail.com}
\affiliation{Kazan Federal University, Kazan, Russia}
\affiliation{Udmurt Federal Research Center of the Ural Branch of the Russian Academy of Sciences, Izhevsk, Russia}

\author{Anatolii~V.~Mokshin} \email{anatolii.mokshin@mail.ru}
\affiliation{Kazan Federal University, Kazan, Russia}
\affiliation{Udmurt Federal Research Center of the Ural Branch of the Russian Academy of Sciences, Izhevsk, Russia}

\begin{abstract} 
Amorphous metallic foams are prospective materials due to unique combination of their mechanical
and energy-absorption properties. In the present work, atomistic dynamics simulations are performed
under isobaric conditions with the pressure $p = 1.0$ atm in order to study how cooling with extremely high rates ($5\times10^{13}$--$5\times10^{14}$ K/s) affects the formation of pores in amorphous titanium nickelide. For equilibrium liquid phase, vaporization temperature $T_{b}$ and the equation of states in the form of $\rho(T)$ are determined. It is found that the porosity of this amorphous solid does not depend on cooling at such high rates, whereas the pore morphology depends on the magnitude of the cooling rate. The obtained results will be in demand in study of mechanical properties of amorphous metallic foams with a nanoporous structure.
\end{abstract}


\maketitle

\section{Introduction}

Nickel-titanium (NiTi) shape-memory alloy is a widely used functional material for many applications
due to its unique mechanical properties, excellent corrosion resistance, and biocompatibility. The phase transformations in NiTi, in particular, austenite-martensite conversions, are well studied~\cite{ref_1,ref_2,ref_3}. One of the significant achievements is the synthesis of NiTi-based
amorphous metallic foams (AMFs) with micron- or nano-sized pores, with low density in combination
with high strength, high surface area, and open porosity~\cite{ref_4}. The main application of AMFs with
micron-sized pores is related with production of bone implants~\cite{ref_5,ref_6,ref_7}, whereas AMFs with nanoporous structure are demanded, for example, in production of energy-storage elements~\cite{ref_8,ref_9}. Moreover, AMFs are widely used as functional materials, where the pores
morphology (closed-cell or open-cell) plays a crucial role~\cite{ref_10,ref_11}. The porous structure may consist of interconnected and/or spatially separated cells, and the presence of such the cells determines the mechanical properties of AMF and, eventually, the fields of its application. Thus, it is important to develop the methods for producing AMFs with a required degree of porosity and necessary features of pore morphology. 

NiTi alloy has the high melting temperature $T_m = 1583$~K. The production of NiTi AMF is carried out mainly using the powder-metallurgy methods~\cite{ref_12,ref_13,ref_14}. Another method of AMF production is generating the bubble-like cells in the melt at a temperature above
the melting (liquidus) temperature with subsequent vitrification of this melt~\cite{ref_11,ref_12,ref_13,ref_14,ref_15}. Upon rapid solidification of the melt, these bubbles form a porous structure of the amorphous material. At relatively low applied cooling rates, this method allows to obtain AMF with a low degree of porosity. In the present work, based on atomistic dynamics simulations we show that the morphology of the pores in NiTi-based AMF can be controlled by means of applied cooling rates.

\section{Atomistic dynamics simulations of amorphous NiTi with nanoporous structure}

We perform atomistic dynamics simulations of Ni$_{50}$Ti$_{50}$ alloy. The interaction energy between atoms is defined through the modified embedded-atom method (MEAM) interatomic potential. This MEAM potential was proposed by Ko et al. for NiTi-based alloys~\cite{ref_16,ref_17}. 
\begin{figure}[ht!]
	\centering\includegraphics[width=1.0\linewidth]{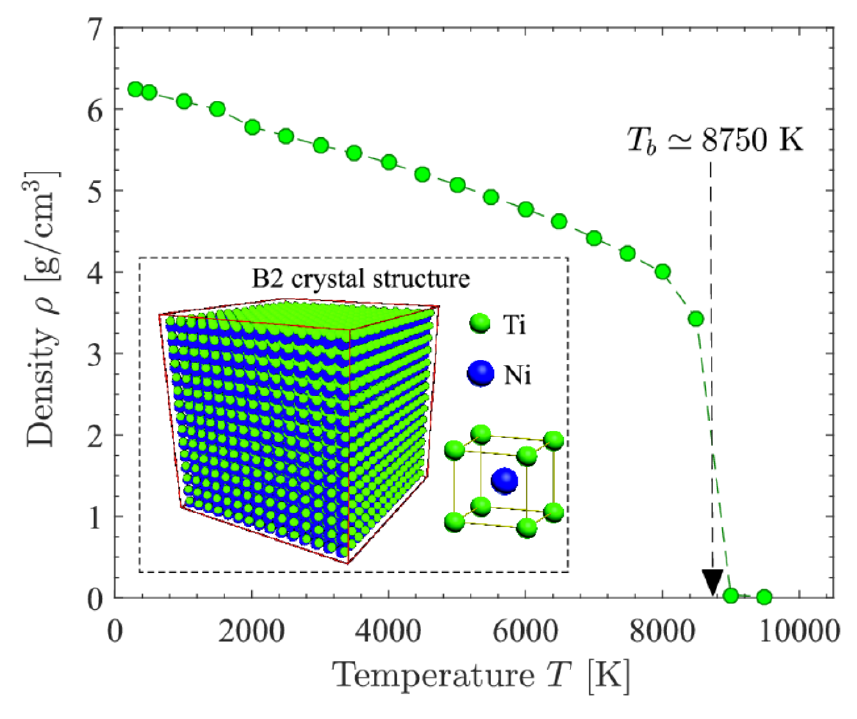}
	\caption{Equation of state, $\rho(T)$ of equilibrium melt Ni$_{50}$Ti$_{50}$ evaluated from atomistic dynamics simulations. Inset: fragment of the B2 crystal structure for Ni$_{50}$Ti$_{50}$ with $6859$ atoms.}
	\label{fig_1}
\end{figure}
\begin{figure*}[ht]
	\centering\includegraphics[width=0.65\linewidth]{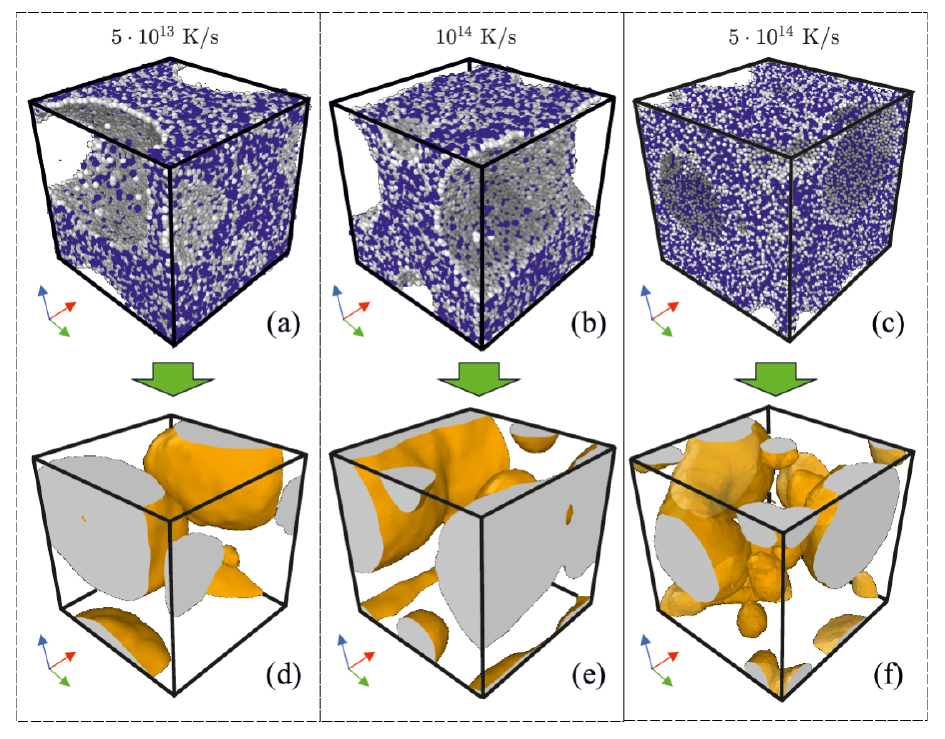}
	\caption{Obtained at various cooling rates, (a–c) 3D snapshots of amorphous NiTi with porous structure at the temperature $300$~K, where the dark and light spheres are Ni and Ti atoms, respectively; (d–f) 3D maps of the inner surface of the pores.}
	\label{fig_2}
\end{figure*} 
The simulation cell with applied periodic boundary conditions in all the directions contains $109744$ atoms ($54872$ atoms of Ni and the same number of Ti atoms). Temperature and pressure are controlled by the Nose–Hoover thermostat and barostat with the damping parameters $Q_T = 100\Delta t$ and $Q_P = 5000\Delta t$, respectively, where the time step is $\Delta t = 1$~fs.

Initially, the simulated system was prepared so that to correspond to the cubic B2 crystalline phase at the temperature $T = 0$~K. Then, this crystalline system was rapidly heated and equilibrated at the ($p$, $T$) states with the pressure $p = 1$~atm and with various temperatures $T$ from the range of $300$--$9500$~K. At each considered ($p$, $T$) state, the system was brought to equilibrium over the time interval of $1$~ns. The mass density of Ni$_{50}$Ti$_{50}$, evaluated from simulation data, is given in Fig.~\ref{fig_1} as function of temperature. 

As seen from Fig.~\ref{fig_1}, at temperatures above $8000$~K the density of the system decreases rapidly, and at the temperature $T_b \approx 8750$~K the ``liquid–vapor'' phase transition is observed. Since we are aiming to generate a porous amorphous solid, we need to start the cooling procedure of a melt with the lowest density. As has been found, the equilibrium liquid sample Ni$_{50}$Ti$_{50}$ with the temperature $T = 8500$~K and the density $\rho = 3.44$~g/cm$^3$ is the most suitable for our purpose. This sample was cooled independently with three various cooling rates. Namely, applying to the sample a rapid isobaric cooling with the cooling rates $5 \times10^{13}$, $10^{14}$, and $5\times 10^{14}$~K/s to the states with the temperature $T = 300$~K, we obtained the three samples of the porous amorphous Ni$_{50}$Ti$_{50}$ alloy. It is important to note that ultrafast cooling with such very high cooling rates was recently carried out experimentally for nanometersized samples~\cite{ref_18}.

\section{Effect of cooling rate on porosity and pores morphology}

Remarkable feature of vitrification by ultrafast cooling is that such cooling does not provide enough
time for the system to form a dense homogeneous structure due to slow and uneven thermal compression~\cite{ref_19,ref_20}. As a result, during a rapid isobaric cooling, pores form in the system. Fig.~\ref{fig_2} shows the snapshots of amorphous Ni$_{50}$Ti$_{50}$ alloy with a nanoporous structure obtained by cooling the melt at three different rates.

Note that the supercooling level of the system at the considered temperature $T = 300$~K is ($T_m – T)/T_m \approx 0.8$, and a porous amorphous phase is stable at such deep supercooling levels (see Figs.~\ref{fig_2}a--\ref{fig_2}c). 

The coefficient of porosity of the system is estimated using the well-known relation~\cite{ref_21}
\begin{equation}\label{eq_1}
\phi =\left(1-\frac{\rho}{\rho_{0}}\right)\times100\%,
\end{equation}
where $\rho$ is the mass density of porous amorphous Ni$_{50}$Ti$_{50}$ alloy, and $\rho_{0} = 6.24$~g/cm$^{3}$ is the mass density of bulk amorphous Ni$_{50}$Ti$_{50}$ without pores at the temperature $T = 300$~K. The parameter $\phi$ can take values from $0$ to $100$\%, with $\phi = 0$\% corresponding to a sample without pores and $\phi = 100$\% corresponding to the case of vacuum (the pore size equals the system size). We have found that the mass density of the Ni$_{50}$Ti$_{50}$ melt increases from $\rho = 3.44$~g/cm$^{3}$ to $\rho = (4.3 \pm 0.15)$~g/cm$^{3}$ during the applied quench procedure from the state with the temperature $8500$~K to the solid state with the temperature $300$~K. It is also found that the generated porous amorphous alloy has the porosity $\rho = (31 \pm 3)$\%, and the value of the coefficient $\phi$ is independent on cooling with such high rates. Note that materials with such a level of porosity are widely used in production of membranes and electrodes for Li-based batteries with high-energy density, where the cell structures with meso- and micro-porosity allow to achieve optimal electrochemical characteristics of the materials~\cite{ref_22,ref_23,ref_24}. 

For analysis of pore morphology, we have constructed the maps of the inner surface of the pores for all the three samples shown in Figs.~\ref{fig_2}a--\ref{fig_2}c. These maps are presented in Figs.~\ref{fig_2}d--\ref{fig_2}f. As seen from Figs.~\ref{fig_2}d--\ref{fig_2}f, the porous structure of the amorphous alloy appears due to the empty cells of spherical shape uniformly distributed throughout the system. Linear sizes of these cells range from $1$ to $10$~nm. It is important to note that there is correlation between the cell concentration and the cooling rate taken in the cooling protocol to generate the porous amorphous alloy. Namely, at vitrification with the cooling rates greater than $10^{14}$~K/s, the amorphous sample is generated, in which the pores coalesce and form hollow ramified tunnels permeating the whole system (see Figs.~\ref{fig_2}e and~\ref{fig_2}f). At cooling rates less than $10^{14}$~K/s, the lower concentration of pores in the amorphous sample is observed, where the pores are mainly isolated. This is clearly seen in Fig.~\ref{fig_2}d, where the geometry of the pores is shown for the case of the porous amorphous alloy generated by cooling at the rate of $5 \times 10^{13}$~K/s.

\section{Conclusions}

In the present work, we have shown that the porous amorphous Ni$_{50}$Ti$_{50}$ alloy can be generated by ultrafast cooling of a low-density melt. The main results of this study can be summarized as follows:

(i) By means of isobaric atomistic dynamics simulations ($p = 1$~atm) with the MEAM potential suggested
by Ko et al. [16], the equation of state of equilibrium melt Ni$_{50}$Ti$_{50}$ $\rho(T)$ is determined, for the temperature range of $300$--$9500$~K.

(ii) Amorphous Ni$_{50}$Ti$_{50}$ alloy with nanoporous structure can be generated by means of ultrafast isobaric cooling, and the generated amorphous samples are characterized by the coefficient of porosity $\phi = (31\pm3)$\%.

(iii) It is shown that the morphology of the pores depends on magnitude of the cooling rate, by means of
which the amorphous alloy was generated. 

(iv) It is found that the amorphous material with isolated pores and the material with percolated porous
structures can be generated due to isobaric ultrafast cooling with various cooling rates.

\begin{acknowledgments}
\noindent This work is supported by the Russian Science Foundation (project no.~19-12-00022).
\end{acknowledgments}

\end{document}